# A Discussion on the Algorithm Design of Electrical Impedance Tomography  for Biomedical Applications

**Mingyong Zhou & Hongyu Zhu**

School of Computer Science and Communication  Engineering
GXUST
Liuzhou,  China
Zed6641@hotmail.com

**Abstract.**

   In this paper, we present a discussion on the algorithms design of Electrical Impedance Tomography (EIT) for biomedical applications. Based on the Maxwell differential equations and the derived the finite element(FE) linear equations, we first investigate the possibility to estimate the matrix that contains the impedance values based on Singular Value Decomposition(SVD) approximations. Secondly based on the biomedical properties we further explore the possibility to recover the impedance values uniquely by injecting various different types of currents with multi-frequency. Injecting various types of multi-frequency currents lead to a set of different measured voltages configurations, thus enhance the possibility of uniquely recovering the impedance values in a stable way under the assumption that the biological cells respond to the different type of injecting currents in a different way.

   By converting the Maxwell differential equations into linear equations by Finite Elements(FE) method, we are able to focus on the discussions based on the linear algebra method.  We also explore some insights into the biological cells' electrical properties so that we can make use of  the biological cell's electrical properties to make the numerical algorithm design more stable and robust.

# 1 Introduction

   Electrical Impedance Tomography (EIT) draws the attention of researchers in biomedical engineering and applied mathematics for its low cost and convenient approaches using the well-know electrical method[1][2][3][5]. From the hardware design including the injecting various currents types to the imaging algorithms development invoking mathematical methods,  fruitful research results of EIT appear in recent years in applied mathematics, industrial areas and biomedical engineering. In this paper, however we focus our discussions on the area of biomedical imaging applications. Imaging in medical application is more complex and challenging because of the complex tissues and cells structures in human bodies.

In this paper, we focus our discussions on the algorithms design as well as the implications of the electrical properties of tissues and cells on the algorithm design. First, we outline the EIT problem as Maxwell differential equations and address the forward problem and inverse problem (imaging problem). Secondly, based on the the Maxwell differential equations and the derived the finite element(FE) linear equations, we investigate the possibility to estimate the matrix that contains the impedance values based on Singular Value Decomposition(SVD) approximations. It is shown that based on the SVD method the solution to the EIT problem is not unique without a prior information on the impedance values . Last but not the least, based on the biomedical properties we further explore the possibility to recover the impedance values uniquely by injecting various different types of currents with multi-frequency. Injecting various types of multi-frequency currents lead to a set of different measured voltages configurations, thus enhance the possibility of uniquely recovering the impedance values in a numerically stable way under the assumption that the biological cells respond to the different type of injecting currents in a different way. It is worth noting that we convert the whole non-linear EIT problem into a linear problem with a very larger and very high dimension based on Finite Element(FE) method, thus allowing one to approach the EIT problem using linear algebra method especially for the biomedical applications.

We devote to our discussions on the implications of the electrical properties on the algorithm design in the last part of this paper but not the least as is declared because this part is particularly important in the applications of biomedical engineering by taking into account the electrical properties of tissues and cells, that is, tissues and cells of human bodies respond to the different frequencies of injecting currents in a different way. We can make use of such electrical properties to ensure that the algorithm design is more stable and robust but unique simultaneously.

## 2 Maxwell differential equations and EIT problem

Inside of magnetic field, magnetic field effect can be ignored due to the quite low permeability in biological tissue. According to the Maxwell theory and Ohm law, we can get the description like this:

$$- \rho^{-1} \nabla \phi = \vec{J} \quad (1.1)$$

In this equation, $\rho$ represents the distribution function of impedance, $\phi$ means potential distribution, J is the function of boundary electric current density.

Due to there is no electric current through interior of biological tissue, so we get this:

$$\nabla \cdot \vec{J} = 0 \qquad (1.2)$$

Combined (1-1) and (1-2), we can get equation :

$$\nabla \cdot \rho^{-1} \nabla \phi = 0 \quad (1.3)$$

This partial differential equation should satisfy the Dirichlet boundary condition

$$\phi \mid_{\partial\Omega} = V \qquad (1.4)$$

In that, $\mathbf{\Omega}$ is the area where the object in. For the electrodes with inject current, Neumann boundary condition is :

$$\rho^{-1} \frac{\partial \phi}{\partial \vec{n}} \bigg| = \vec{J} \quad (1.5)$$

The equations above constitute the mathematical model of EIT problem[5].

Given the impedance distributions and the injecting currents J , it is required to calculate the voltages on the surface. This is termed as a forward problem. For forward problem of EIT, we now summarize the procedure of finite element (FE) method to formulate a set of linear equations for the Maxwell differential equations represented by equations(1-1) to (1-5).

## 3 Finite Element(FE) method to formulate the EIT forward problem

For a 3D bounded by $\Omega$, we can divide the 3D bounded space $\Omega$ with the following triangle with K,M,N,L and coordinates represented by （$x_k,y_k,z_k$）, $(x_M,y_M,z_M)$，$(x_N,y_N,z_N)$，$(x_L,y_L,z_L)$.

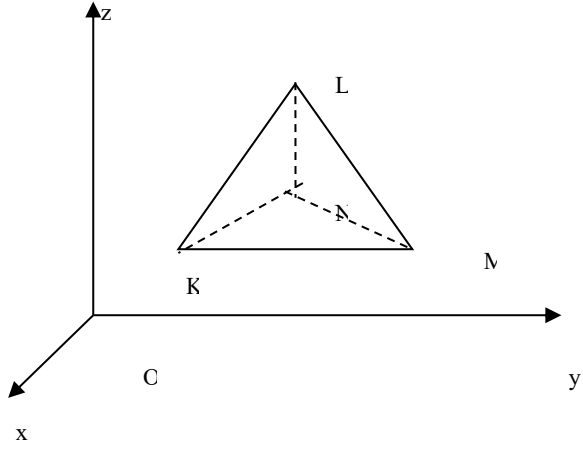

**Figure2.1: Finite Element Triangle**

The derivation of the Finite Element(FE) method is too tedious but is a well-known and well used method in mechanics engineering. By diving the bounded space Ω with small triangles and enough numbers of triangles, for the ith injection of currents we can formulate a linear equations as follows:

$$S_i \Phi_i = F_i \quad （2.1）$$

where $F_i$ is a vector that is unrelated to impedance values $\rho$ and is related only to injecting currents，$\Phi_i$ is a vector formulated as the voltages measured on each element and $S_i$ is a matrix formulated by the impedance values $\rho$ of each triangle elements. Now the EIT inverse problem becomes to estimate the values of Matrix $S_i$, given the measured voltages vector $\Phi_i$ and vector $F_i$ that is formulated as the injecting currents. In the following section, we present a subspace fitting approach based on SVD method to estimate the Matrix $S_i$ and discuss the uniqueness of the estimation. We particularly take a Higher Order Statistics(HOS) approach to remove the Gaussian noise.

Regarding the equation (2.1), in the following section we provide a subspace fitting approach based on SVD method that is originally proposed by Petre Stoica for Signal Processing[4 ].

## 4 SVD method to solve the EIT inverse problem

We now re-formulate the equation (2.1) in another formation as follows by taking into the account of measurement noise denoted by n(t).

$$\vec{y}_{M \times 1}(t) = A_{M \times d} \cdot \vec{x}(t) + \vec{n}(t) \tag{3.1}$$

The following assumptions are made:

A), matrix A 's dimension is M x d , where d<M,

B), the columns of matrix A is linearly independent,

C), $\vec{n}(t)$ is a white or colored Gaussian noise by EIT measurements,

and $E\{\vec{y}\,\vec{y}^T y_i\}$ as the calculated cumulant matrix from the calculations of

Higher orders statistics of measured data exist for all i's, where i=1,2...M. . The problem is to find all solutions of matrix A that minimizes the result of L in the following equation (3.2) .

$$L = \| \vec{y}_{M \times 1}(t) - A \cdot \vec{x}(t) \|_2 \tag{3.2}$$

We summarize the algorithm based on SVD here as follows:

Denote the SVD of any cumulant matrix $E\{\vec{y}\,\vec{y}^T y_i\}$ as follows:

$$(U \mid V)\begin{pmatrix} \sum & 0 \\ 0 & 0 \end{pmatrix}\begin{pmatrix} S & T \\ G & T \end{pmatrix} = U \sum S^T_{dxM} \tag{3.3}$$

By P. Stoica in [4], one obtains $E\{\vec{y}\,\vec{y}^T y_i\}$ ,as the follows

$$AD_i A^T G = U \sum S^T G = 0 \tag{3.4}$$

Where $D_i = \sum_{j=1}^{d} A_{ij} E\{\vec{x}\,\vec{x}^T x_j\}, (i = 1, 2..., M)$ . Under the assumptions of A),B) and C), the approximation problem becomes

$$\min_A \min_C \| A - SC \|_F \tag{3.5}$$

In reference [4], the optimal solutions are derived as the $d^2$ eigenvectors that correspond to the least minimum $d^2$ eigenvalues of the following matrix Q

$$Q \overset{\Delta}{=} \left\{ I - (\underset{M \times d}{S} \otimes \underset{d \times d}{I}) \left[ (S \otimes I)^u (S \otimes I) \right]^{-1} (S \otimes I)^H \right\} \quad （3.6）$$

Where $\otimes$ denotes Kronecker product.

## 5 A simple experiment to demonstrate SVD method

We carried a simple example experiment to demonstrate the method based on SVD method. First the algorithm is outlined as follows. Note that 2$^{nd}$ statistics are used for the simple experiment and measurement noise is not taken into account in the case.

1), $\overset{\to}{y}_{M \times 1}(t) = A_{M \times d} \cdot \overset{\to}{x}_{d \times 1}(t) \quad (M \geq d)$        (4.1)

Where $\overset{\to}{y}(t)$ is a vector formulated by the measured voltages.

2), $E\left\{ \overset{\to}{y}(t)\ \overset{\to}{y}^T(t) \right\} \quad Y_{M \times M}$        (4.2)

Where $Y_{M \times M}$ is a correlation matrix that is formulated based on a 2$^{nd}$ order statistics.

3) , Denote SVD of cumulant matrix by：

$$Y_{M \times M} = U_{M \times d} \sum_{d \times d} S_{d \times M}^T \quad (4.3)$$

4),calculate the matrix Q as follows:

$$Q \overset{\Delta}{=} \left\{ I - (\underset{M \times d}{S} \otimes \underset{d \times d}{I}) \left[ (S \otimes I)^T (S \otimes I) \right]^{-1} (S \otimes I)^T \right\} \quad (4.4)$$

The eigenvectors of Q with dimension of $Md \times 1$ thus corresponding to the least minimum eigenvalues of Q form all possible solutions to vector $\overset{\to}{A}_{Md \times 1}$, and $\overset{\to}{A}_{Md \times 1}$ is a column vector with dimension $Md \times 1$ derived from matrix A.

One should note that the solutions to matrix A are not unique by SVD method as proposed as the above.

A simple experiment was carried as demonstrated in Figure 5.1, where the circles are placed with resistors A-E and only one resistor is placed in the center of circle

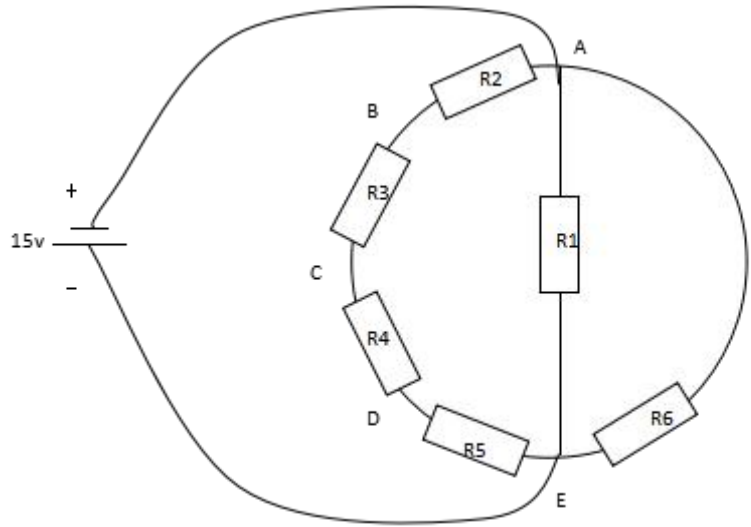

**Figure 5.1: A EIT experiment based on SVD method**

The 4 voltages at point B,C,D,E with reference to point A are measured as a 4 x 1 data vector, that is M=4. Algorithm's steps 1)-5) are used for the estimation in which d=3 is used. The derived Q in this experiment is thus a matrix with rows of 12 and columns of 9 (12 x 9) .

The estimation solutions to matrix A are derived as the following nine 4 x3 matrix results after the calculations. We note that in all 9 matrix solutions there is only one non-zero value 1.0 that is actually corresponding to the only one resistor in the center of the circle.

$$
\begin{bmatrix} 0 & 0 & 0 & 1 \\ 0 & 0 & 0 & 0 \\ 0 & 0 & 0 & 0 \end{bmatrix}
\begin{bmatrix} 0 & 0 & 0 & 0 \\ 1 & 0 & 0 & 0 \\ 0 & 0 & 0 & 0 \end{bmatrix}
\begin{bmatrix} 0 & 0 & 0 & 0 \\ 0 & 1 & 0 & 0 \\ 0 & 0 & 0 & 0 \end{bmatrix}
\begin{bmatrix} 0 & 0 & 0 & 0 \\ 0 & 0 & 1 & 0 \\ 0 & 0 & 0 & 0 \end{bmatrix}
\begin{bmatrix} 0 & 0 & 0 & 0 \\ 0 & 0 & 0 & 1 \\ 0 & 0 & 0 & 0 \end{bmatrix}
$$

$$
\begin{bmatrix} 0 & 0 & 0 & 0 \\ 0 & 0 & 0 & 0 \\ 1 & 0 & 0 & 0 \end{bmatrix}
\begin{bmatrix} 1 & 0 & 0 & 0 \\ 0 & 0 & 0 & 0 \\ 0 & 0 & 0 & 0 \end{bmatrix}
\begin{bmatrix} 0 & 1 & 0 & 0 \\ 0 & 0 & 0 & 0 \\ 0 & 0 & 0 & 0 \end{bmatrix}
\begin{bmatrix} 0 & 0 & 1 & 0 \\ 0 & 0 & 0 & 0 \\ 0 & 0 & 0 & 0 \end{bmatrix}
$$

# 6  Further discussion on the implications of electrical properties on cells and tissues of animals and human bodies

The results in section 5 are based on one current injection which is not a unique result as one observes in the experiment. We now focus on the discussion regarding equation (2.1) that is derived from Finite Element(FE) method. Following many reference in biology and biomedical experiments, e,g,in [6], some cells and tissues are made of water and special materials which up to present is still unknown to human beings but demonstrate many electrical properties resembling that of capacitors as well as resistors at least. In particular the cells and tissues in different parts of the animal and human bodies vary with different circumstances such as temperatures, activity and injecting direct or alternative current frequencies. Such electrical properties provide one an opportunity to estimate the impedance through different aspects as long as the response of cells and tissues to different circumstances are not simply a "linear" response.

For equation (2.1) denoted by the above formulation for the ith Alternative Current(AC) injection, suppose we inject the AC currents with different frequencies, the value $\Phi_i's$ should vary at least not strictly linearly by the cells and tissues electrical properties.For each AC current injections with different frequencies,

$$S_1\Phi_1 = F_1 \text{ , } S_2\Phi_2 = F_2 \text{ ...}$$ we assume matrix $S_1, S_2...$ etc remain the same that is formulated from the impedance distributions (and this assumption is rational because the impedance itself does not change much during the measurement time). Now the equations becomes $S \Phi_1 = F_1$ , $S \Phi_2 = F_2$ ... etc. With this assumption and observation, we can have a new formulation with a much higher dimension as follows:

$$S[\Phi_1, \Phi_2...\Phi_N] = [F_1, F_{2...}, F_N]$$

(6.1)

Where N is the injection current iteration number with different frequencies and temperatures. $[\Phi_1, \Phi_2...\Phi_N]$ is a non-singular matrix formulated by the measured voltages with different AC frequencies and temperatures.

With the proper selection of frequencies and temperatures, it is possible to obtain a more stable and unique solution to the EIT inverse problem. Further clinic empirical data is to be collected to prove the assumption and approaches however.

# 7 Conclusions

Although this paper presents a discussion on the EIT algorithm development based on Finite Element(FE) method and linear algebra methods, and no general conclusions can be made , this paper explores methods based on linear approximations for EIT inverse problem for biomedical engineering applications. At least 3 conclusions can be made , 1), linear approximations are possible based on Finite Element(FE) method and SVD approach, 2), it is not possible for one to uniquely determine the impedance values under only one single current injection measurement based on this SVD approach, and  3), by using electrical properties of cells and tissues of animals and human bodies, it is possible to obtain a unique and stable solution. But this depends upon the  clinic data to support the electrical properties of cells and tissues. We present a discussion and a linear method in the last section on how to make use of the electrical properties in animal and human bodies by carrying out multiple measurements with different AC frequencies and temperatures.

## 8 Acknowledgement